\documentclass[preprint2]{aastex}
\usepackage{natbib,epsf,amsmath}
\shorttitle{Hot Plasma in Supernova Remnants}
\shortauthors{Tilley \& Balsara}
\begin{document}
\title{Anisotropic Thermal Conduction in Supernova Remnants: Relevance to Hot Gas Filling Factors in the Magnetized ISM}
\author{David A. Tilley\altaffilmark{1} and Dinshaw S. Balsara$^2$}
\altaffiltext{1}{dtilley@nd.edu$\;\;\;^2$dbalsara@nd.edu\\Department of Physics, University of Notre Dame, Notre Dame, Indiana, USA 46556}
\begin{abstract}
We explore the importance of anisotropic thermal conduction in the evolution of supernova remnants via numerical simulations.  The mean temperature of the bubble of hot gas is decreased by a factor of $\sim 3$ compared to simulations without thermal conduction, together with an increase in the mean density of hot gas by a similar factor.  Thus, thermal conduction greatly reduces the volume of hot gas produced over the life of the remnant.   This underscores the importance of thermal conduction in estimating the hot gas filling fraction and emissivities in high-stage ions in Galactic and proto-galactic ISMs.  
\end{abstract}
\section{Introduction}
\begin{table}
\caption{Initial conditions of the ISM in our simulations. $\rho$ is the density in amu $\mathrm{cm}^{-3}$; T is the temperature in K; B is the magnetic field in $\mu$G; and L is the grid size in parsecs.\label{table_ic}}
\begin{tabular}{lcccc}
Run & $\rho$ & T & B & L\\
\hline 
L0 & 0.7 & 8000 & 0.0 & 300\\
L1 & 0.7 & 8000 & 3.0 & 300\\
H0 & 5.0 & 10000 & 0.0 & 200\\
H1 & 5.0 & 10000 & 6.0 & 200\\ \hline
\end{tabular}
\end{table}
\begin{figure}[t]
\plottwo{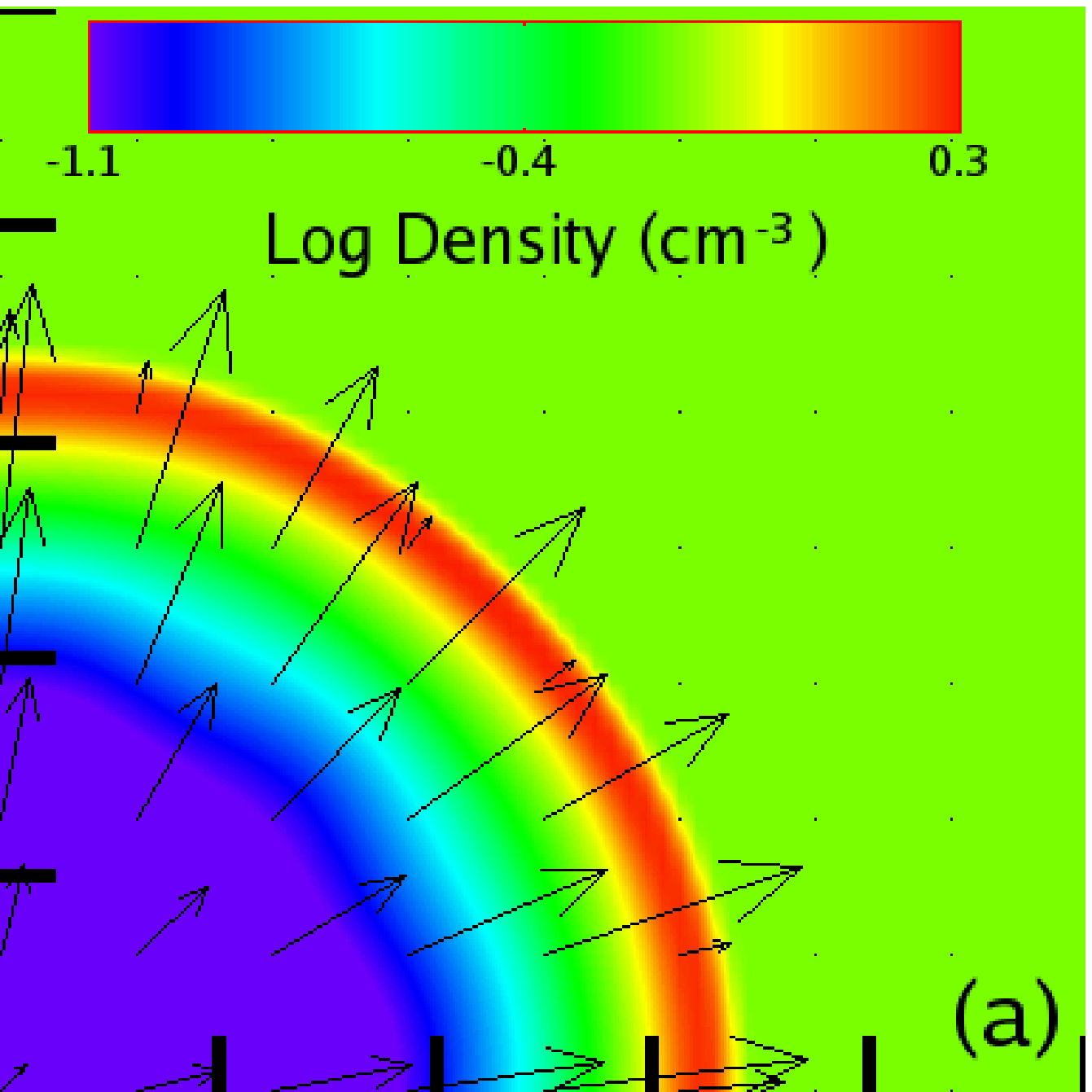}{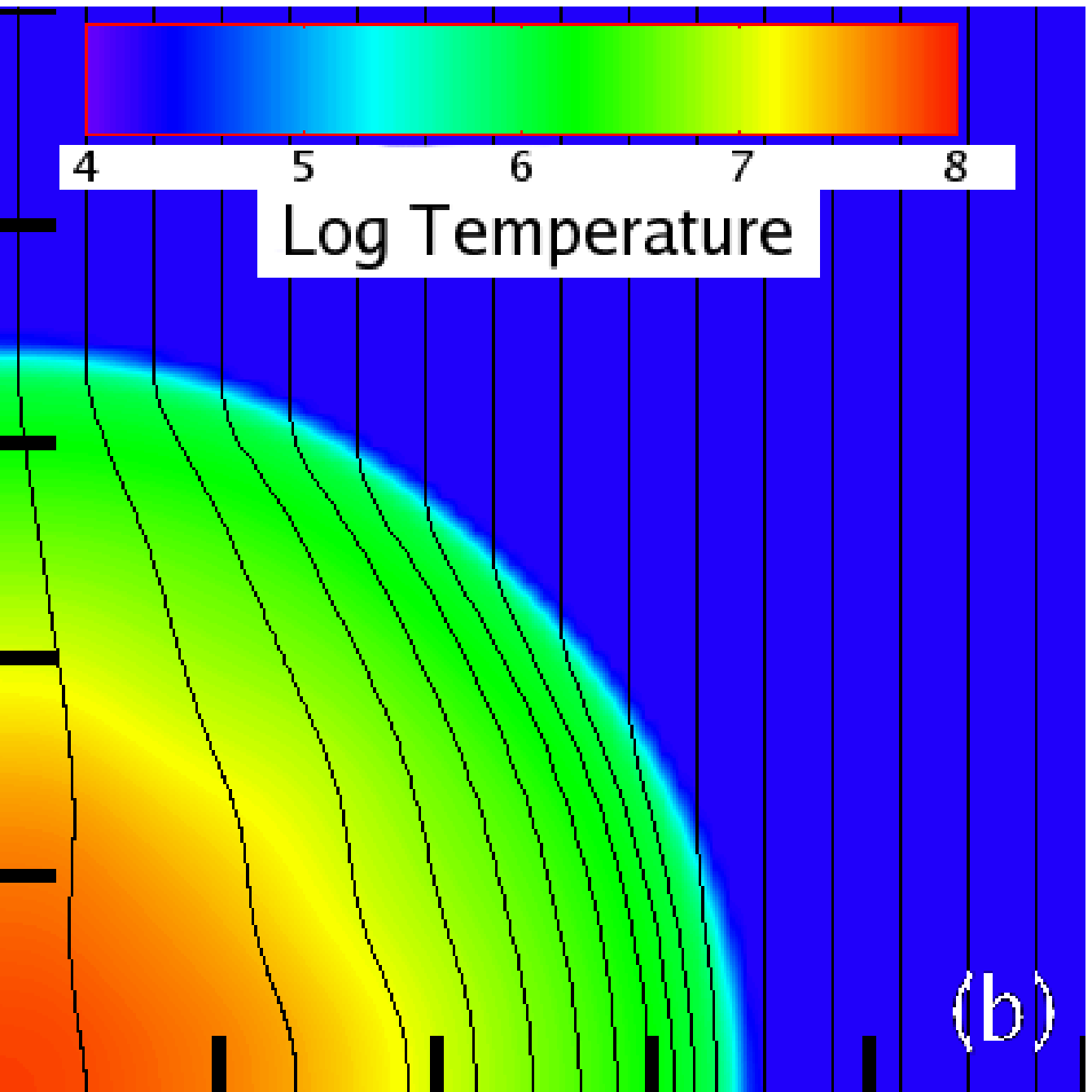}\\
\plottwo{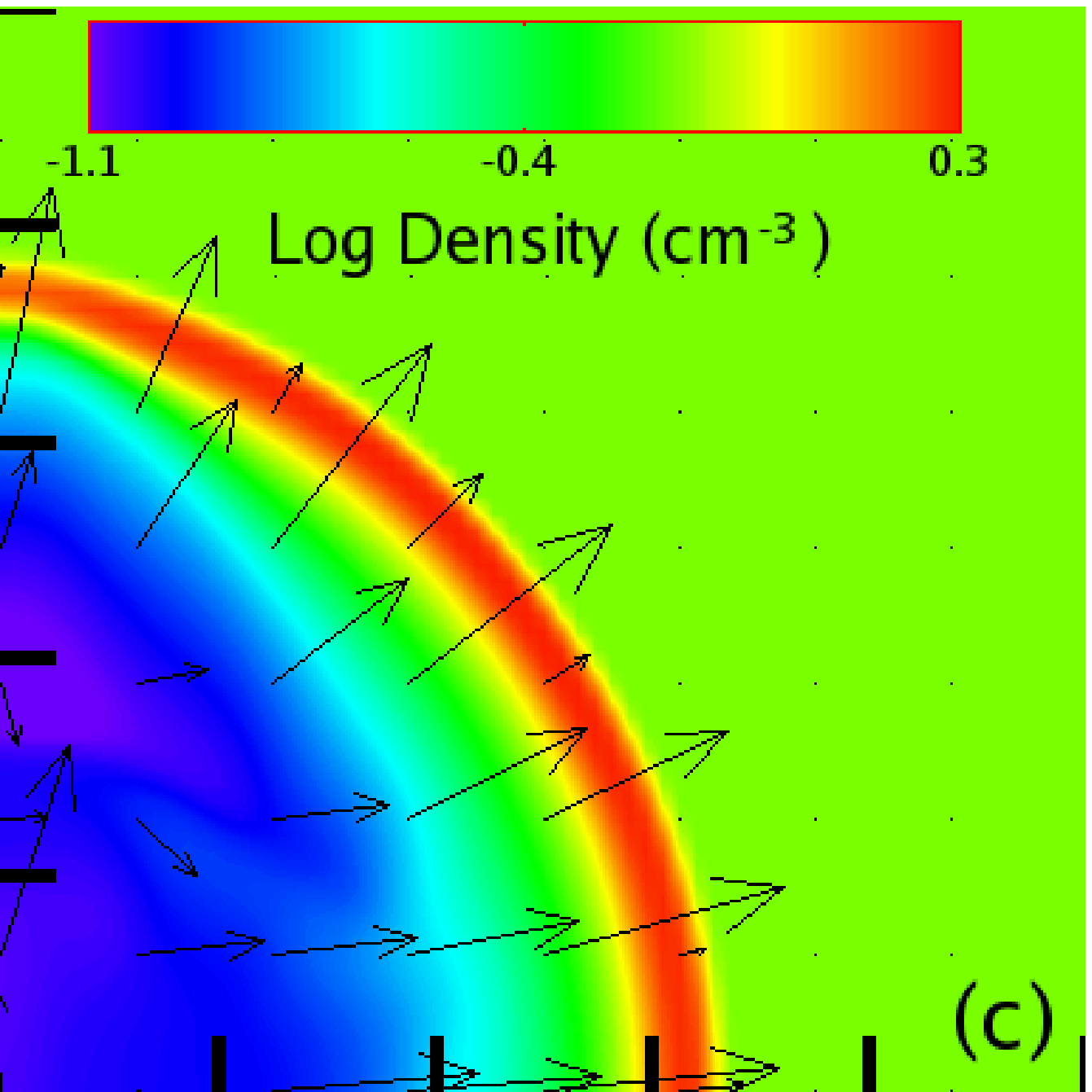}{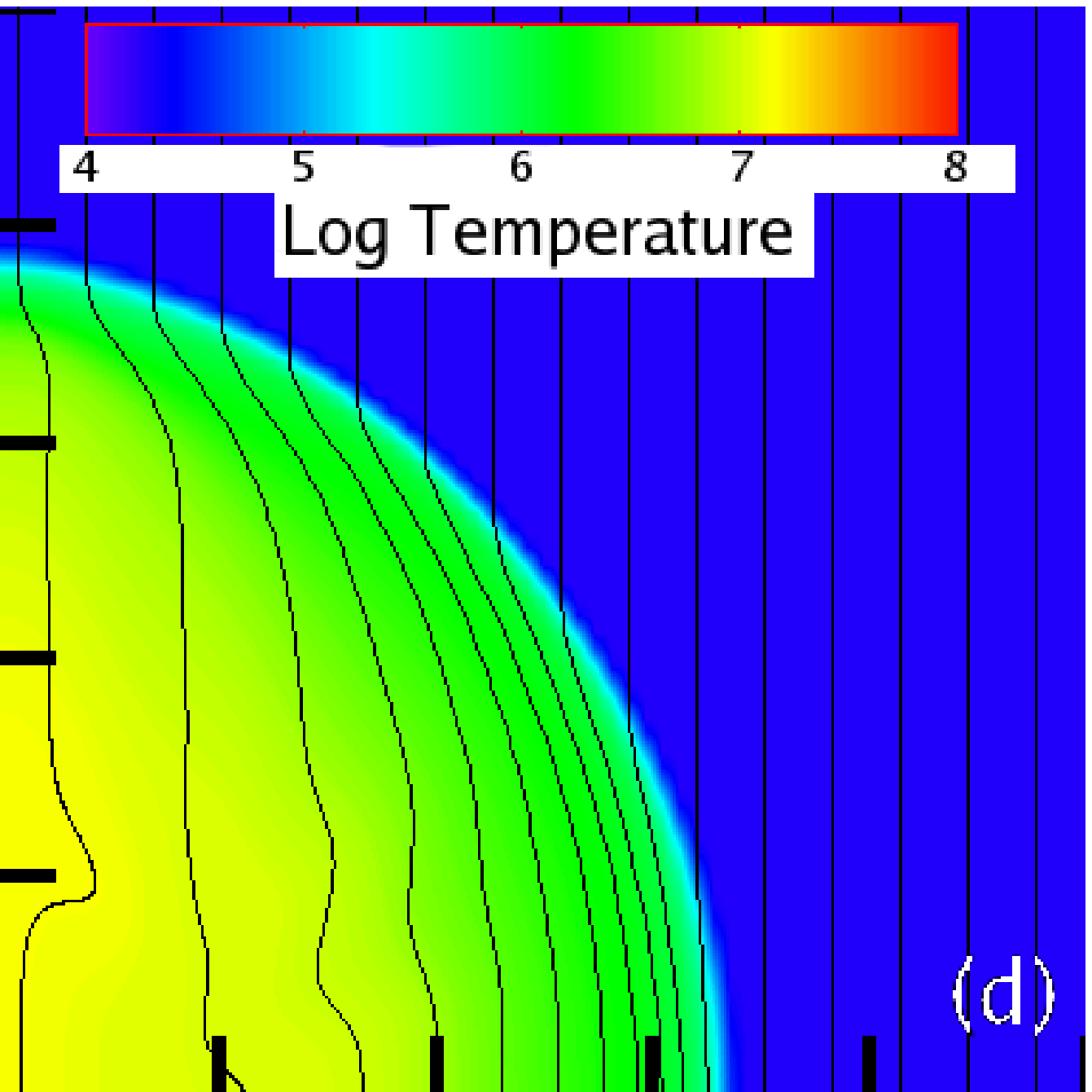}
\vspace{-2mm}\caption{Log density overlaid with velocity vectors (a and c) and log temperature overlaid with magnetic field lines (b and d) for Run L1 at 60 kyrs.  The upper panels exclude thermal conduction, while the lower panels include thermal conduction.  The tick marks are 10 pc apart.\label{fig_dt}}
\end{figure}

Supernovae play an important role in driving the thermal evolution of the interstellar medium (ISM henceforth).  The interiors of supernova remnants (SNR henceforth) are filled with extremely hot plasma, with temperatures of $10^6-10^7\;\mathrm{K}$.  However, the interaction of these remnants with the ISM is not well understood.  Observational clues to the nature of this interaction are made possible by the spectacular maps and spectroscopy of x-ray emission provided by the Chandra space telescope \citep{hwang_etal04}, and by X-ray spectra from Chandra and FUSE of high-stage ions like O \textsc{vi} \citep{bowen_etal04,oegerle_etal05, savage_lehner06}.

The time evolution of the volume of the hot gas produced by a SNR has important implications for the thermal evolution of the multiphase ISM, as described by \citet{field_goldsmith_habing69},\citet{cox_smith74}, and \citet{mckee_ostriker77} (MO77 henceforth).  A comprehensive study has been performed by \citet{slavin_cox92,slavin_cox93} (SC92, SC93 henceforth) in one dimension.  MO77 had estimated that the hot gas would occupy 60-70\% of the volume of the ISM.  The models of SC92 and SC93 found that the porosity of hot gas is reduced by thermal conduction relative to the prediction of the MO77 model, suggesting that the warm phase, and not the hot phase, might be the pervasive phase of the ISM.  The volume filling factor as well as the amount of gas that shines in high-stage ions such as O \textsc{vi} and O \textsc{viii} has recently been measured by FUSE and Chandra \citep{oegerle_etal05, yao_wang05, savage_lehner06}.  Such data has made it possible to subject the MO77 model to strong observational tests.  Consistent with these tests, numerical modelling of the phase structure of the gas has also been advanced by \citet{maclow_balsara_kim_avillez05} and \citet{avillez_breitschwerdt05}.  All such simulations have ignored the role of thermal conduction.  In this paper we examine the role of thermal conduction on the evolution of the 4-volume of the hot gas in the ISM.  This question has immense bearing on the phase structure of the ISM and has assumed immense topical importance in view of the observations and simulations cited above.  

SC92 and SC93 used a simplified representation of the magnetic field in their early calculations of the evolution of SNRs with thermal conduction, and we improve on that here.  \citet{velazquez_martinell_raga_giacani04} include the anisotropy in the thermal conduction introduced by the magnetic field, but do not include the Lorentz force of the magnetic field on the gas.  The inclusion of tension forces from magnetic fields makes the evolution of the SNR anisotropic \citep{balsara_benjamin_cox01}.  Furthermore, in the presence of a magnetic field, thermal conduction itself becomes anisotropic.  Thus, the geometry of the magnetic field cannot be neglected when calculating the thermal heat flux in a multidimensional calculation.

In this Letter we report on the results of a series of MHD simulations on the evolution of SNRs with anisotropic thermal conduction.  We explore a parameter space of initial ISM conditions in density, temperature, and magnetic field strength (Section \ref{section_numerical}).  In Section \ref{section_mmr} we analyze the effects the environment has on the evolution of the remnant.  We examine the porosity of hot gas in these different environments in Section \ref{section_4volume}.  Section \ref{section_conclusion} provides some conclusions.

\section{Numerical Setup}\label{section_numerical}
\begin{figure*}[t]
\plotone{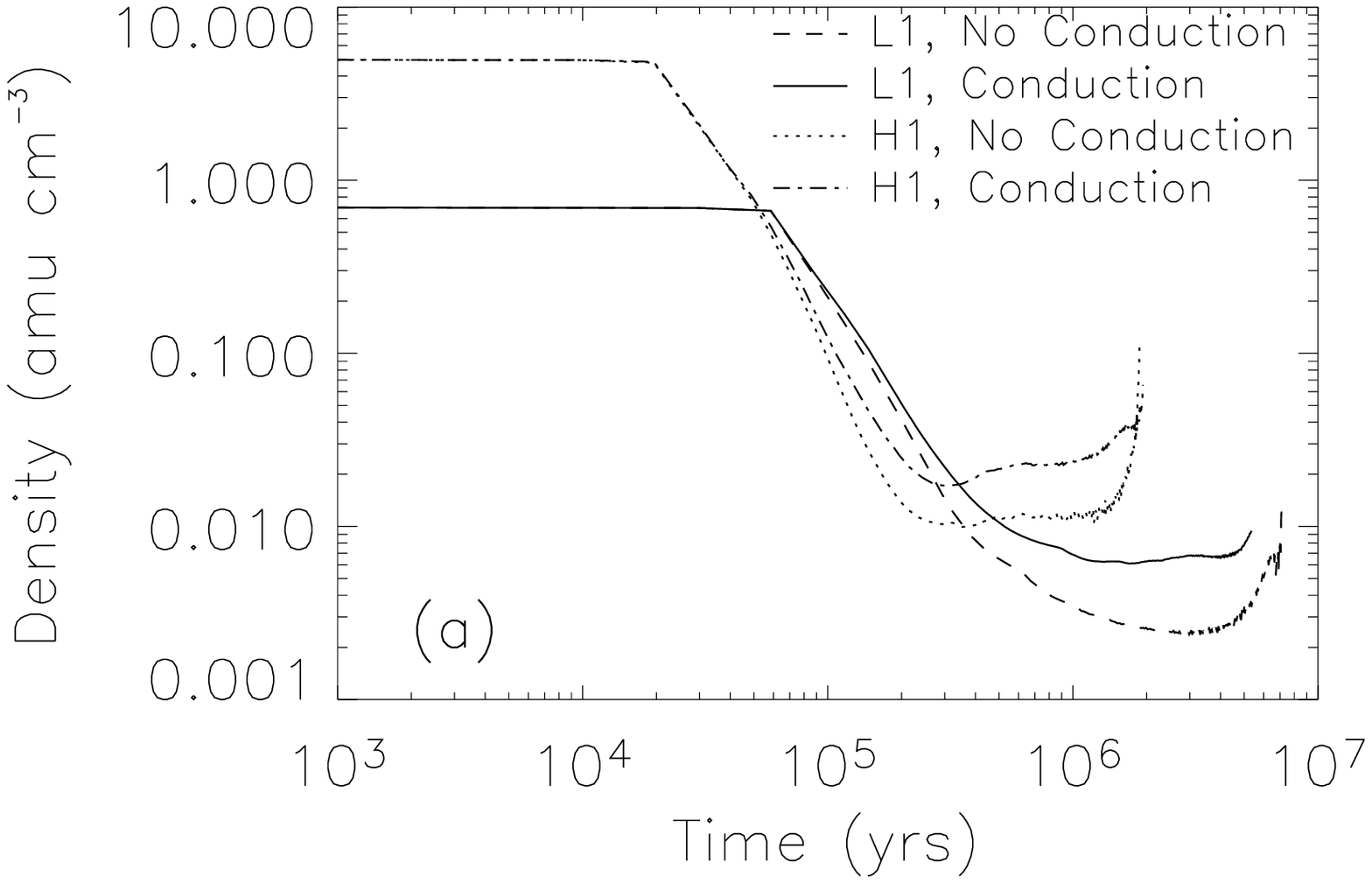}\plotone{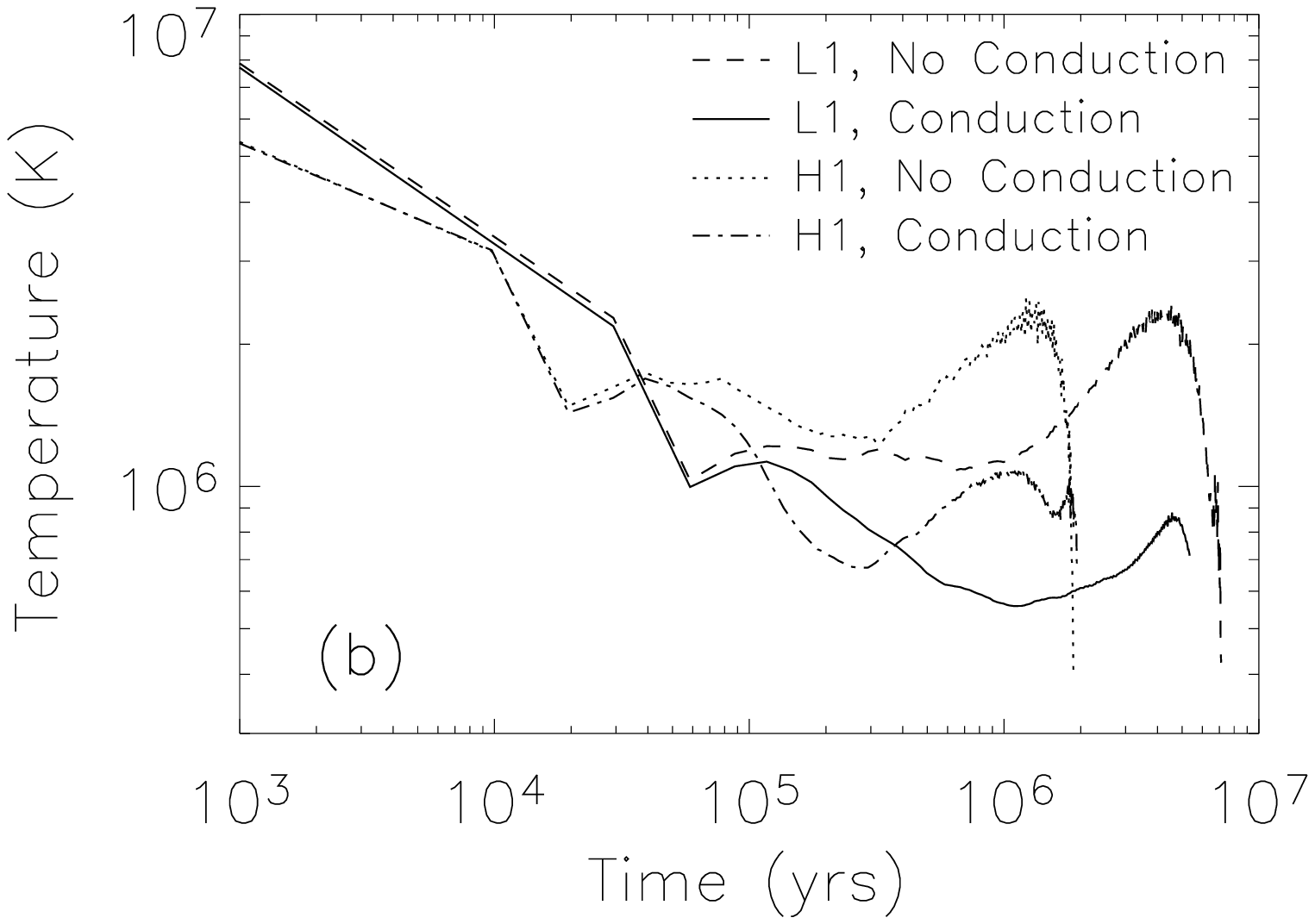}
\vspace{-4mm}\caption{Time-evolution of the mean density (a) and mass-weighted temperature (b) of the gas in the hot bubble of the SNR. This plot compares the results of runs L1 and H1 with and without thermal conduction present.\label{fig_meandenstemp}}
\end{figure*}
\begin{figure*}[tp]
\plotone{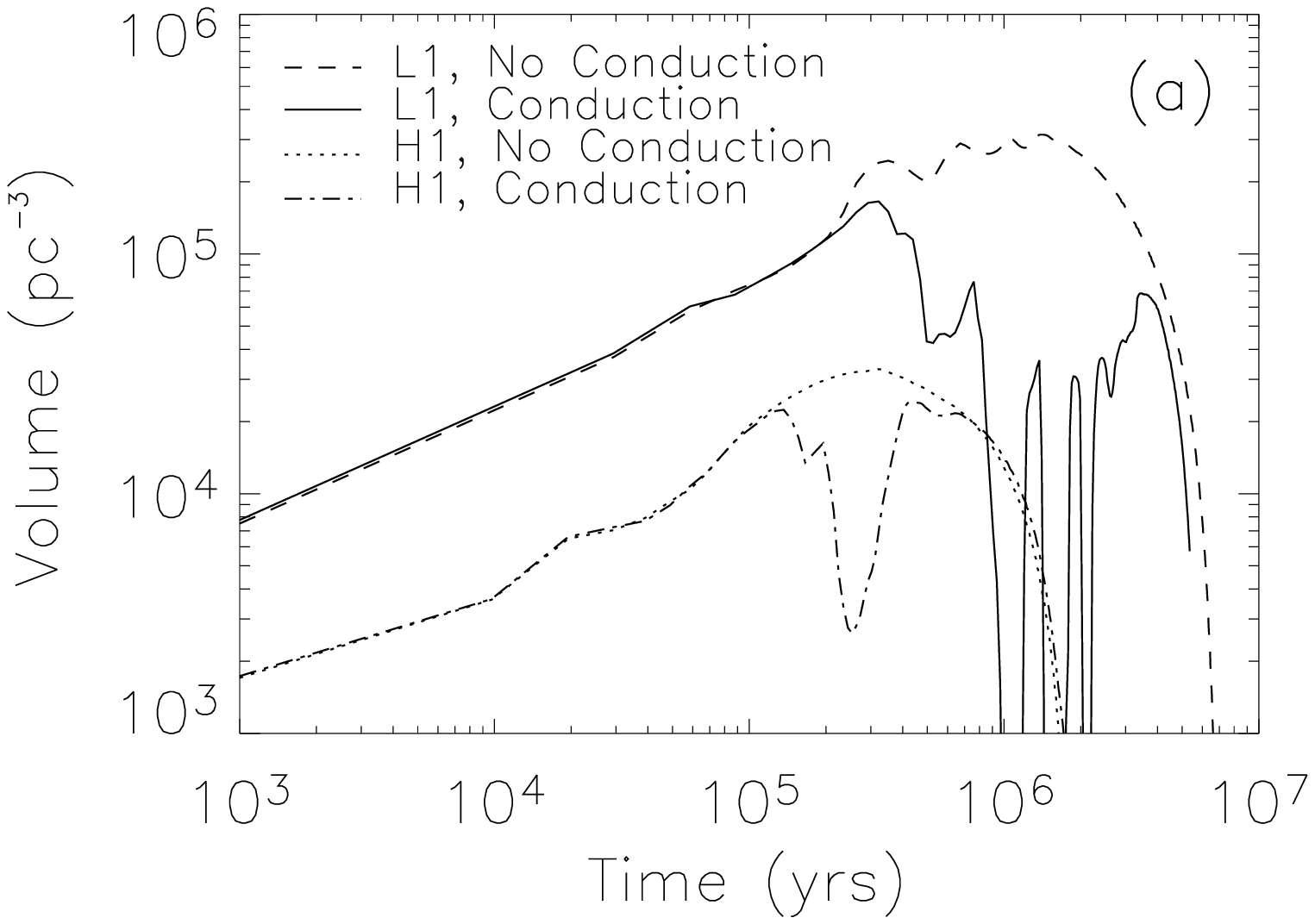}\plotone{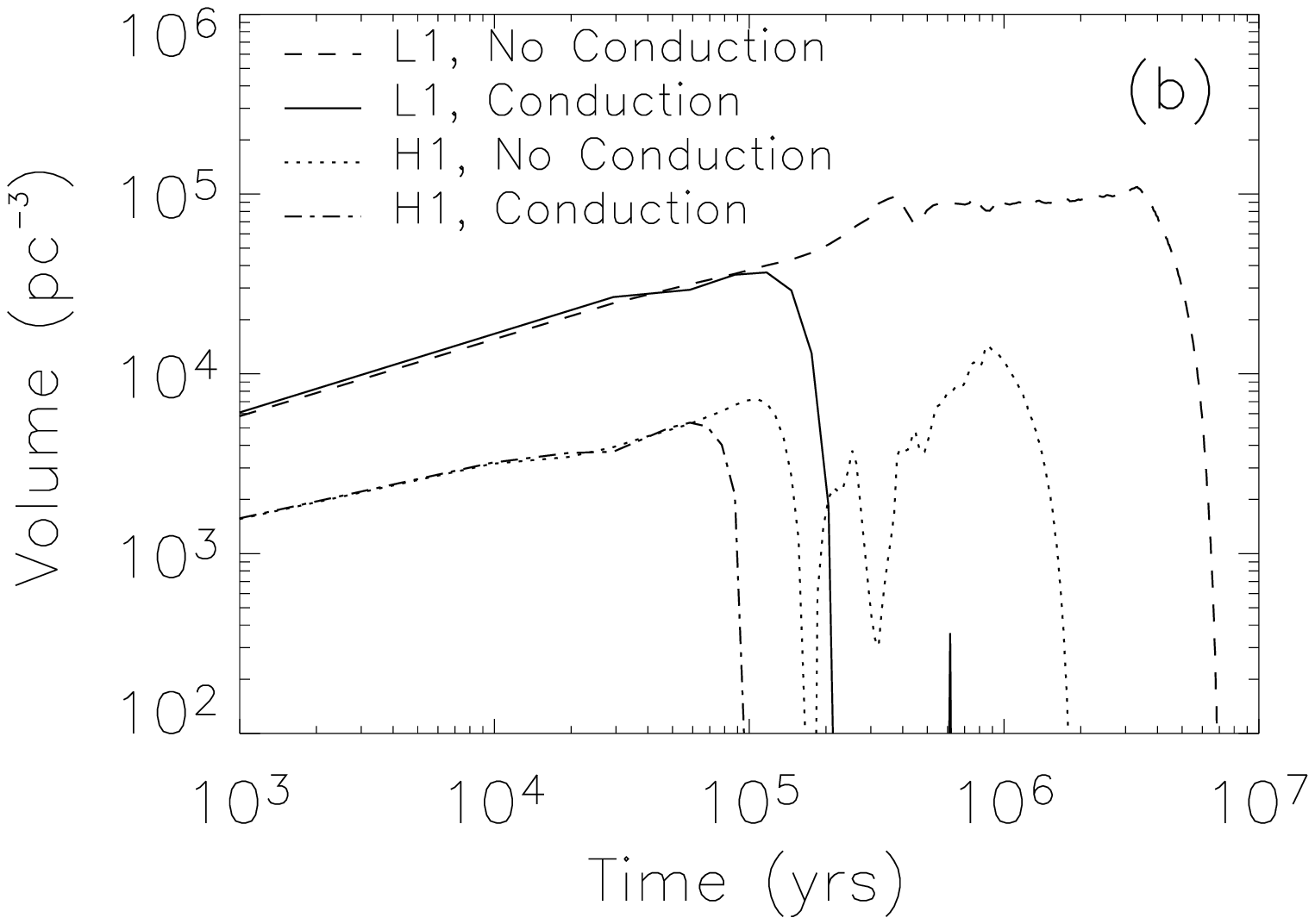}
\vspace{-4mm}\caption{Time-evolution of the volume of hot gas in Run L1 and H1, with and without thermal conduction.  (a) Volume of gas with temperature greater than $7.9\times 10^5$ K.  (b) Volume of gas with temperature greater than $2.2\times 10^6$ K.\label{fig_hotbubble}}
\end{figure*}

Our simulations are performed using the \textsc{riemann} code; see \citet{balsara04} and references therein.  We discuss the details of our treatment of anisotropic thermal conduction with tests and applications to SNRs in an accompanying paper (Balsara et al., in preparation).  We draw on \citet{cowie_mckee77a,balbus86}; SC92 for our formulation of the classical and saturated thermal conduction fluxes.  We utilize a flux limiter to transition from the classical flux to the saturated flux (SC92).  The transition from classical to saturated conduction causes the flux to change character from parabolic to hyperbolic \citep{balbus86}.  We account for this with a Newton-Krylov multigrid method that correctly incorporates the anisotropic structure of the thermal conduction operator in the presence of a magnetic field.

  We solve the MHD equations in cylindrical geometry, using a mesh that is 384x384 uniformly spaced zones in the R-z plane.  We use the cooling function of \citet{macdonald_bailey81} for radiative cooling, with heating balancing cooling in the quiescent ISM.

  We choose our initial conditions to span a range of interesting ISM densities and temperatures.  The values of the density, temperature and magnetic field in the initially quiescent medium are given in Table \ref{table_ic}.  Runs L0 and H0 label SNRs propagating into low and high density ISMs without magnetic fields.  Runs L1 and H1 catalogue similar propagation into ISMs with a magnetic field of 3 $\mu$G.  We initialize our SNRs with $10^{51}$ ergs of energy.  The physical size of our grid is chosen to capture the outer shock through the duration of our simulations.

\section{The Structure of SNRs with Thermal Conduction}\label{section_mmr}

Fig. \ref{fig_dt}(a) and \ref{fig_dt}(b) show the logarithmic density with velocity vectors overlaid, and the logarithmic temperature with magnetic fields overlaid, for Run L1 without thermal conduction, at a time of 60 kyr.  Fig. \ref{fig_dt}(c) and \ref{fig_dt}(d) show the same for Run L1, now repeated with thermal conduction.  We notice that the location of the outer shock in the radial direction is unaffected by the inclusion of thermal conduction.  We also observe that at this early epoch the hot gas bubble with thermal conduction is somewhat smaller in Fig. \ref{fig_dt}(c) and \ref{fig_dt}(d), showing that thermal conduction does reduce the size of the hot gas bubble.  This difference becomes even more exaggerated at later times.  We also observe that the mean density in the runs with thermal conduction is higher, and the mean temperature is correspondingly lower.  Fig. \ref{fig_dt}(b) shows that the hot gas bubble has a temperature up to $10^8$ K, while Fig. \ref{fig_dt}(d) shows that the temperature in the bubble when thermal conduction is included is at most $10^7$ K.  Thus we see that the inclusion of thermal conduction predisposes the hot gas bubble to be strongly radiative in soft x-rays, a feature not found in the simulations that exclude thermal conduction.

We show the time evolution of the mean density and temperature of the hot bubble in Fig. \ref{fig_meandenstemp}, along with a comparison with the evolution of density and temperature for the same initial conditions, but without thermal conduction.  The mean densities and temperatures shown in Fig. \ref{fig_meandenstemp} include all gas with temperature $T > 10^5$ K.  We see that the inclusion of thermal conduction produces a half-order of magnitude greater density in each of the SNRs studied here.  Likewise, the temperature decreases by half an order of magnitude.  This is directly related to the fact that thermal conduction exchanges energy between the colder and hotter parts of the remnant.  Because the late-time evolution of remnants is isobaric, the decrease in the bubble's mean temperature is matched by a corresponding increase in its mean density when thermal conduction is included.  The curves terminate after $\sim 2$ Myr (in the high-density ISM) and $\sim 7$ Myr (in the low-density ISM), as is apparent from Fig. \ref{fig_meandenstemp}, due to the collapse of the hot bubble.

\begin{table*}[t]
\caption{Porosity of hot gas, defined as $S_{-13}\int \mathrm{V}(t) \mathrm{d}t$ where $S_{-13}$ is the Galactic supernova rate in units of $10^{-13}\;\mathrm{pc}^{-3}\;\mathrm{yr}^{-1}$ .  V(t) is the volume of hot gas that exceeds a specified temperature.\label{table_4volume}}
\begin{tabular}{c|llll}
Gas & \multicolumn{2}{c}{With conduction} & \multicolumn{2}{c}{Without conduction} \\
Temperature (K) & Run L1 & Run H1 & Run L1 & Run H1 \\
\hline $> 2.2\times 10^6$ & $5.11\times 10^{-4} S_{-13}$ & $3.63\times 10^{-5} S_{-13}$ & $4.26\times 10^{-2} S_{-13}$ & $1.02\times 10^{-3} S_{-13}$ \\
$> 7.9\times 10^5$ & $2.09\times 10^{-2} S_{-13}$ & $2.12\times 10^{-3} S_{-13}$ & $9.00\times 10^{-2} S_{-13}$ & $2.66\times 10^{-3} S_{-13}$ \\
$> 3\times 10^5$ & $9.99\times 10^{-2} S_{-13}$ & $3.05\times 10^{-3} S_{-13}$ & $1.03\times 10^{-1} S_{-13}$ & $2.82\times 10^{-3} S_{-13}$
\end{tabular}
\end{table*}
\section{Four-Volume of Hot Gas}\label{section_4volume}

The amount of hot gas deposited into the ISM controls its subsequent thermal evolution.  As the remnant matures, cooling and thermal conduction eventually cause the hot bubble to cease expanding and begin to collapse.  This is illustrated in Fig. \ref{fig_hotbubble}, which plots the volume of gas greater than $7.9\times 10^5\;\mathrm{K}$ and $2.2\times 10^6\;\mathrm{K}$ as a function of time.  The relevance of these temperature for high-stage ions is explained in the next paragraph.  The area under these curves provides us a measure of the 4-volume at temperatures exceeding the above-mentioned values.  Comparing the dashed line and solid line in Fig. \ref{fig_hotbubble}(a) shows that the inclusion of thermal conduction dramatically reduces the amount of hot plasma in the bubble with temperatures exceeding $7.9\times 10^5$ K after $3\times 10^5$ years.  After 1 Myr we see a small emergence of hot gas owing to the fact that the interior is reheated by shocks in the collapsing bubble.  Comparing the dotted and dot-dashed lines in Fig. \ref{fig_hotbubble}(a) confirms that a similar scenario prevails for SNRs going off in denser media.  Fig. \ref{fig_hotbubble}(b) is even more illustrative as it shows that there is a dramatic difference in the 4-volume of gas hotter than $2.2\times 10^6$ K when thermal conduction is included.  In this case, the plasma is hardly reheated to these temperatures by the shocks driven into the collapsing hot gas bubble.

The porosity of hot gas is defined in Table \ref{table_4volume} and follows the formulation of SC93.  For situations where our runs have similar parameters to those in SC93, we cross-checked our measures of the 4-volume with theirs and found good consistency.  Table \ref{table_4volume} shows the 4-volumes for different runs for temperatures greater than $3\times 10^5$ K, $7.9\times 10^5$ K, and $2.2\times 10^6$ K.  These temperatures correspond to the temperatures at which the emission is dominated by O \textsc{vi} (measured by FUSE) and O \textsc{vii} and O \textsc{viii} (measured by Chandra).  The amount of gas injected into the ISM at hotter temperatures is considerably reduced compared to the $3\times 10^5$ K gas.  Thus, the conclusions of SC93, that the filling factor of hot gas from SNRs in the ISM is considerably reduced from the standard predictions of the MO77 model, are strengthened when anisotropic thermal conduction is added to models of SNRs.  Table \ref{table_4volume} shows that thermal conduction reduces the gas that could emit in O \textsc{vi} by a small factor but the gas that could emit in O \textsc{vii} and O \textsc{viii} by large factors.  These results translate to protogalactic ISMs where they may even influence the physics of early galaxy formation.

It has been speculated \citep{cho_etal03,avillez_breitschwerdt05,balsara_kim05} that high rates of turbulent diffusion might make thermal conduction irrelevant in global studies of the ISM.  \citet{balsara_kim05} find that very strongly driven turbulence has a coherence time of 0.8 Myr.  We anticipate Galactic turbulence to be much more weakly driven, with a larger coherence time.  Dispersal of the hot gas bubble would only take place after several coherence times.  Such times would be comparable to the lifetime of the hot gas bubble catalogued in Table \ref{table_4volume}.  As a result, we do not expect interstellar turbulence to completely erase the structure of the hot gas bubble before it collapses.  We therefore expect our conclusions to be robust in the presence of interstellar turbulence. 

\section{Conclusion}\label{section_conclusion}

Anisotropic thermal conduction decreases the average temperature of hot gas by half an order of magnitude over a period of a few million years, and increase the mean density by a similar amount.  It also changes the 4-volumes of x-ray emitting gas.  The emission in high-stage ions is also significantly altered by the inclusion of thermal conduction.  This result is important not just for simulations of our present Galaxy, but also in cosmological simulations of proto-galaxies.

The authors warmly acknowledge discussions with R. Benjamin and C. Howk.  DSB acknowledges support via NSF grants AST-005569-001 and NSF-PFC grant PHY02-16783.  The simulations were performed at UND and NCSA.  This work has made use of the NASA Astrophysics Data System abstract database.
\bibliographystyle{apj}
\bibliography{ref}
\end{document}